\documentclass[aps,prb,reprint,superscriptaddress,amsmath,amssymb]{revtex4-1}
\usepackage{xcolor} % required for coloring
\usepackage{graphicx}% Include figure files
\usepackage{dcolumn}% Align table columns on decimal point
\usepackage{bm}% bold math
\usepackage[utf8]{inputenc}
\usepackage[T1]{fontenc}
\usepackage{mathptmx}
\usepackage{etoolbox}
\usepackage{gensymb}
\usepackage[colorlinks=true,linkcolor=black,citecolor=black,urlcolor=black]{hyperref}

\newcommand{\sfrac}[2]          { {\textstyle \frac{#1}{#2}} }
\providecommand{\ket}[2][\vert] { #1{#2} \rangle }

\begin{document}

\title[]{Topological Hall Response from Canted Antiferromagnetic Order in $d$-Electron Kagome Systems}
    \author{$^{*}$Waquar Ahmed}
    \author{Steffen Schäfer}
    \author{Pierre Lombardo}
    \author{Roland Hayn}

    \author{Imam Makhfudz$^{\dagger}$}

    \affiliation{IM2NP, UMR CNRS 7334, Aix-Marseille Université, 13013 Marseille, France}
    \begin{abstract}
        In a two-dimensional kagome monolayer, a nontrivial intrinsic Berry curvature may arise in the $d$-electron system from the interaction with a non-collinear spin order induced by an underlying antiferromagnetic exchange.  This opens the route for a quantum anomalous Hall effect in the multi-orbital system, even without an external magnetic field, explicit spin-orbit coupling or relativistic effects. For spin orders with an out-of-plane component, the scalar spin chirality is finite, and the integration of the Berry curvature over the Brillouin zone may yield integer Hall conductivities in units of $e^2/h$.  For a Fermi level within a nontrivial gap, the canted configuration offers, at least in principle, the possibility of a maximal Chern number, $C=\pm 5$.  Candidate materials are considered in this paper. In existing materials, the electron hopping is generally highly anisotropic, leading to a quantum anomalous Hall effect with smaller Chern numbers. A topological phase transition between Hall plateaus of opposite $C$ can be driven by flipping the out-of-plane component of the spin order, alluding to the potential of this system to applications in quantum information.   
    \end{abstract}
    \maketitle
    
% %%%%%%%%%%%%%%%%%%%%%%%%%%%%%%%%%%%%%%%%%%%%%%%%%%
\section{Introduction}
    The Anomalous Hall Effect (AHE), where an electric current produces a transverse voltage due to time-reversal symmetry breaking without an external magnetic field, has been at the forefront of condensed matter physics, both due to its theoretical interest and its potential applications in dissipationless electronics.\cite{AHErmp}   
    Millennial-era advances in condensed matter physics are marked by the discovery of materials with electronic structures hosting peculiar features, such as Dirac points and flat bands.\cite{Flatband1,Flatband2,Flatband3,CominNatureMaterials0} In the presence of magnetism, such materials may form a frustration-induced quantum spin liquid.\cite{MakhfudzPRB2014,SavaryBalentsRPP} Alternatively, the magnetic order may induce an AHE, which can be driven by extrinsic mechanisms such as impurities or disorder.
    A more fundamental approach, by contrast, aims at an AHE in clean systems, solely relying on phenomena within the material itself, e.g. an internal magnetic texture\cite{Taguchi,YePRL,OnodaNagaosaPRL2,MartinBatista,OhgushiPRB,ZFangScience,CanalsPRB,MacDonaldPRL,BrunoPRL,MakhfudzPujolPRB2015} yielding an effective magnetic field in reciprocal space.\cite{SundaramNiu,Sinitsyn,Haldane2004PRL} This \emph{Berry curvature} may then give rise to the quantum anomalous Hall effect (QAHE),\cite{MacDonaldRMP} characterized by an integer\cite{TKNN,AHErmp} -- the Chern number. 
    It was theoretically predicted to occur in a variety of systems,\cite{Haldane1988PRL,OnodaNagaosaPRL1,KSunPRL2009,QAHEgrapheneAFM,QAHEinGraphene,CXLiuPRL,CXLiuAnnCondMat}, several of which were experimentally confirmed.\cite{Science,ScienceDeng} 
    A recent work of ours shows a QAHE for the $d$-electrons in a two-dimensional kagome lattice with ferromagnetic order. \cite{MakhfudzPRB2024}
      
    A key ingredient of both classical and quantum AHE is the breaking of time-reversal symmetry. In ferromagnets, the symmetry breaking is due to a non-vanishing net magnetization. In antiferromagnets, however, the net magnetization is zero and the magnetic ordering not controllable via an external magnetic field. This has spurred the conventional belief that antiferromagnets are not suitable for applications. This view has been challenged by recent developments, demonstrating that some effects may still be explored, thus giving rise to emerging fields like antiferromagnetic spintronics.\cite{ManchonRMP}
    
    It is well understood that in collinear bipartite antiferromagnets, e.g. on the honeycomb lattice, spin staggering implies a zero net magnetization and a spin-degenerate electronic structure. However, a zero net magnetization may also arise in a less trivial manner by non-collinear ordering, for example on triangular and kagome lattices.\cite{Sachdev1992PRB} 
    On a kagome lattice with antiferromagnetic interactions, the classical AHE has been predicted theoretically from explicit spin-orbit coupling\cite{MacDonaldPRL} and observed experimentally via its anomalously large Hall conductivity.\cite{ScienceLargeAHE,NatureLargeAHE} 
    Canted ferromagnetic and antiferromagnetic orders near collinearity, but with spin chirality, have been predicted to give rise to the chiral Hall effect\cite{ChiralHECP} which was observed experimentally in van-der-Waals materials with non-coplanar spin order.\cite{TakagiNatPhys}

    \begin{figure}
	    \includegraphics[angle=0,origin=c, scale=0.25]{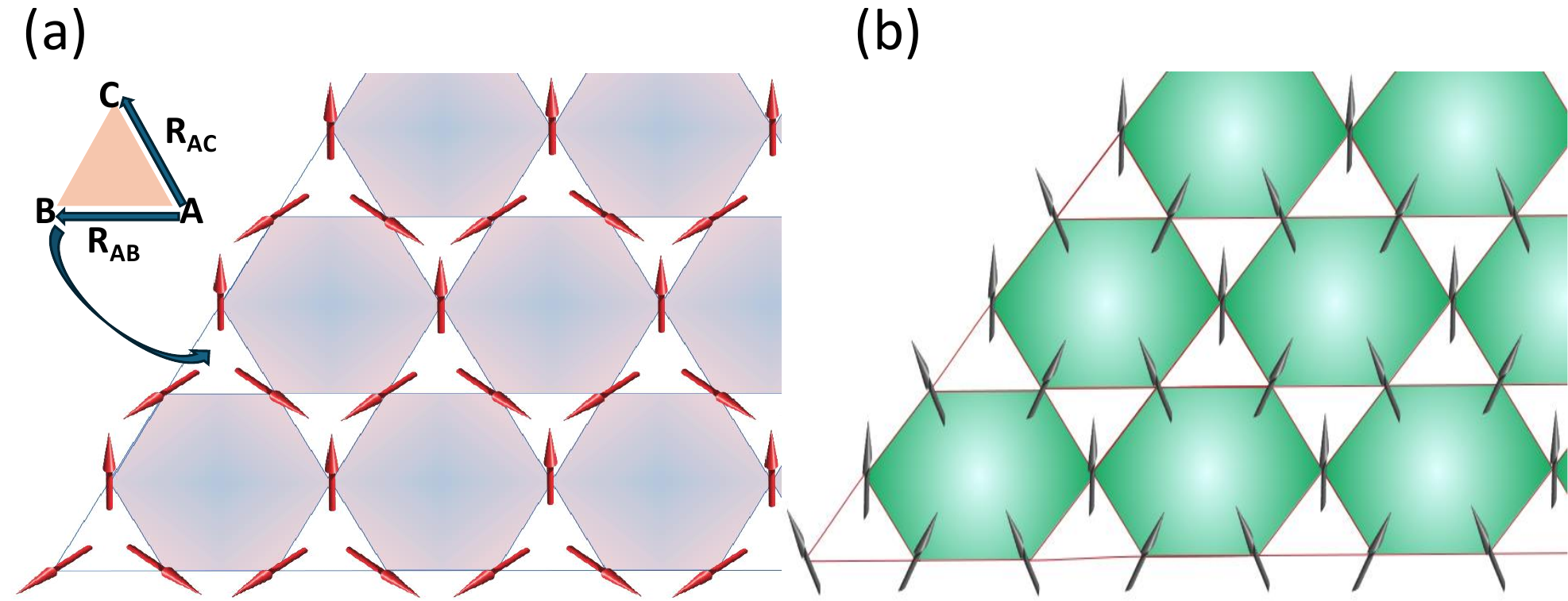}
	    \caption{Two-dimensional kagome lattices with (a) coplanar non-collinear spin order with three sublattices carrying spins at 120° from each other; and (b) non-coplanar spin order by adding an out-of-plane component to the spins.
              In both cases, the spins on the three sublattices are related to each other by $120\degree$ rotations about the center of the white triangle in the unit cell.}
        \label{fig:KagomeLatticeandBZ}
    \end{figure}
    \begin{figure*}
        \includegraphics[width=6.5in]{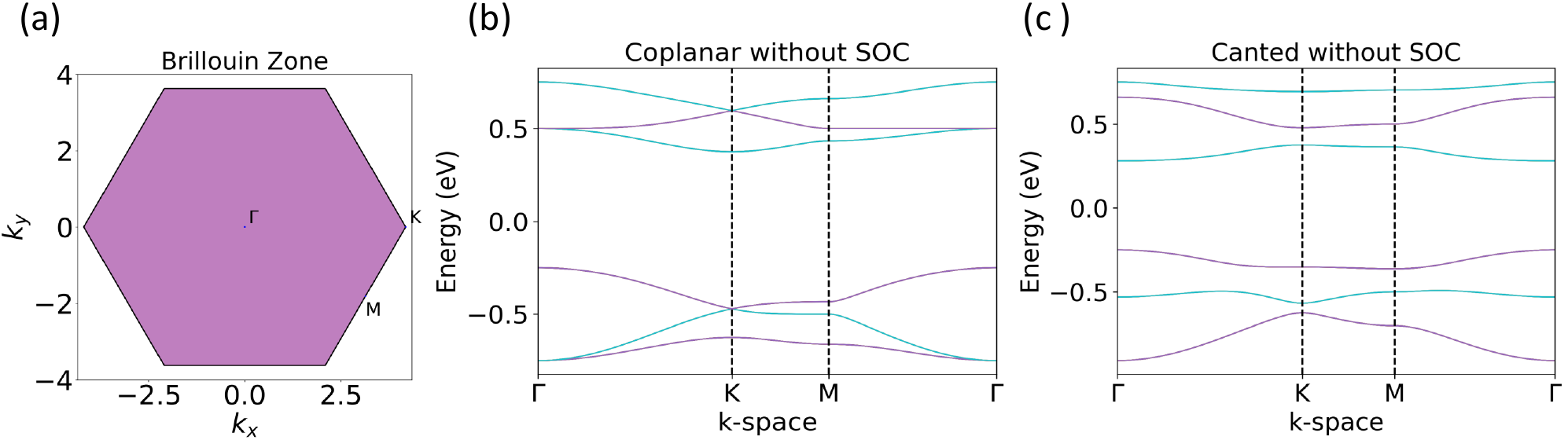}
	      \caption{(a) First Brillouin zone of the 2D kagome lattice, with symmetry points $\Gamma$, $K$, and $M$.
        (b),(c) Band structures along the path $\Gamma-K-M-\Gamma$ for the Hamiltonian (\ref{Hamiltonian}), with isotropic hopping, and spin splitting arising due to non-collinear spin order. (b) Coplanar non-collinear spin order, with a large trivial gap around $E=0$. 
        (c) Non-coplanar spin order, with a large trivial gap and four small nontrivial gaps.
        Parameters: $E_1=E_2=E_3=0\,$eV, $V_{dd\pi}=V_{dd\delta}=V_{dd\sigma}=-0.25\,$eV, and $\mathbf{M}(\mathbf{r})=(M_x(\mathbf{r}),M_y(\mathbf{r}),M_z(\mathbf{r}))=M_s(\sin\theta_{\mathbf{r}}\cos\phi_{\mathbf{r}},\sin\theta_{\mathbf{r}}\sin\phi_{\mathbf{r}},\cos\theta_{\mathbf{r}})$, with $M_s=1\,$eV, azimuthal angles $\phi_{\mathbf{r}}=120^{\circ}$, and polar angles $\theta_{\mathbf{r}}=90^{\circ}$ and $\theta_{\mathbf{r}}=53^{\circ}$, in (b) and (c), respectively. 
       }
        \label{fig:BandStructure}
    \end{figure*} 
    In heterostructured kagome systems, the QAHE due to canted spin order of localized $d$-orbital electrons was studied theoretically for $s$-orbital conduction electrons.\cite{ZhouNanoLett} 
    In transition-metal based compounds such as metal-organic frameworks (MOFs),\cite{YaghiNature, Shaiek} however, the conduction electrons are $d$-type and the fate of the QAHE in such multi-orbital systems is still unclear.  Such compounds are inherently two-dimensional and display narrow energy bands, differing from both porous 3D MOFs and kagome metals which consist of coupled inorganic kagome layers and are usually characterized by wide bands.
    A natural question we want to address in the present article is therefore whether the QAHE can occur in a multi-orbital $d$-electron system with non-collinear spin order but \textit{in the absence} of explicit spin-orbit coupling (SOC) term,  since the orbital degree of freedom might either disrupt or reinforce the QAHE. The coupling of the localized spin texture to the conduction electrons may induce a spin-dependent effective flux in real space, as manifested in complex hopping integrals for the electrons and the spin splitting of the electronic structure, even without net magnetization. Coplanar non-collinear spin order turns out to give no QAHE due to trivial real-space fluxes, unless a transfer-type explicit spin-orbit coupling is introduced.\cite{MacDonaldPRL}

    The topological QAHE requires non-coplanar non-collinear spin order. Here, we demonstrate this theoretically in a multi-orbital system, where all five $d$-orbitals contribute to the Chern number, but each spin sector with opposite sign. The highly degenerate case,  which is promoted by isotropic electron hopping and equal onsite energies for all $d$-orbitals, produces a maximal Chern number of C = $\pm 5$. The opposite signs for opposite spin sectors result from spin--momentum locking. In real materials, by contrast, this key condition is usually not fulfilled, and the observed Chern numbers are generally smaller.  Nevertheless, the QAHE still involves contributions from different orbitals.              
    %\vspace{-1.80em}
    
% %%%%%%%%%%%%%%%%%%%%%%%%%%%%%%%%%%%%%%%%%%%%%%%%%%
\section{Model}
    
    We consider a tight-binding model where the $d$-electrons are subject to orbital-dependent onsite energies (1st term), hopping (2nd term), and couple to the local spin order via a Zeeman exchange (3rd term):
    \begin{align}
        \label{Hamiltonian}
        H = &
        \sum_{\substack{i\\ \alpha\sigma}} E_{\alpha}d^{\dag}_{i,\alpha\sigma}d_{i,\alpha\sigma}
        \,+\,\frac{1}{2}\sum_{\substack{\langle ij\rangle \\ \alpha\alpha'\sigma}}
        \left[t_{ij,\alpha\alpha'}d^{\dag}_{i,\alpha\sigma}d_{j,\alpha'\sigma}+\mathrm{h.c.}\right] \nonumber
        \\
        &-\sum_{\substack{i\\ \alpha\sigma\sigma'}} d^{\dag}_{i,\alpha\sigma}\left(\mathbf{M}_i\cdot\mathbf{s}\right)_{\sigma,\sigma'}d_{i,\alpha\sigma'}
    \end{align}
    where $d^{\dag}_{i,\alpha\sigma}(d_{i,\alpha\sigma})$ are creation (annihilation) operators for $d$-electrons of spin $\sigma$ in the orbital $\alpha$ and on site $i$ as illustrated in Fig.~\ref{fig:KagomeLatticeandBZ}.  Although the onsite energies may be different for all five $d$-orbitals, we consider the special case where only three different values occur, namely $E_1$ for the $d_{z^2}$-orbital, $E_2$ for $d_{xz}$ and $d_{yz}$, and $E_{3}$ for $d_{xy}$ and $d_{x^2-y^2}$, with the remaining degeneracies expressing lattice symmetries.
    The hopping elements $t_{ij,\alpha\alpha'}$ in the second term are linear functions of the Slater-Koster integrals $V_{dd,\tau=\sigma,\pi,\delta}$ between two $d$ orbitals.\cite{Harrison} This makes the $d$-electron hopping orbital and direction dependent, and allows for an exchange between states belonging to different orbitals. The Zeeman coupling in the third term results in a spin splitting due to an exchange with the local spin order.  The corresponding magnetic moments for the sublattices are $\mathbf{M}_{i\in A,B,C}$, and their planar projection is assumed to be at 120° with respect to each other as illustrated in Fig.~\ref{fig:KagomeLatticeandBZ}. The net magnetization $M_z$ is zero for the coplanar case, but nonzero for a non-coplanar or canted spin order due to the out-of-the-plane component of the spins. We will see below that the non-collinear magnetic structure gives rise to a non-relativistic spin splitting. 

% %%%%%%%%%%%%%%%%%%%%%%%%%%%%%%%%%%%%%%%%%%%%%%%%%%
\section{Band structure and non-relativistic spin splitting} 
       
    The first step is to compute the band structure of our Hamiltonian Eq.~(\ref{Hamiltonian}), consisting of electronic bands arising from three sublattices, with 5 orbitals and 2 spin states each. The Fourier transformation under periodic boundary conditions then yields the $\mathbf{k}$-space Hamiltonian.

    The resulting band structures are displayed in Fig.~\ref{fig:BandStructure},
    for both non-collinear but coplanar spin order and non-collinear non-coplanar order. In the present case, the hopping terms have been chosen equal, implying Slater-Koster integrals of maximal symmetry, with $V_{dd\sigma}=V_{dd\pi}=V_{dd\delta}$. This gives the simplest band
    structure, with almost flat bands and linear or quadratic Dirac crossing points.\cite{CominNatureMaterials0} This situation will be called the \textit{'isotropic limit'}. Our model is nevertheless completely general, and readily applicable to real materials, where the Slater-Koster integrals will generally be anisotropic, and may even be of different signs. The results turn out to be qualitatively robust, but differ quantitatively. This affects the maximally attainable quantized Hall conductivity.  We will first focus on the isotropic limit, for the sake of simplicity and analytical tractability. The results apply qualitatively also to real materials discussed in the last sections, but additional features may arise from extremely anisotropic Slater-Koster integrals.  In these sections, we also put forth quantitative predictions that could be tested experimentally.

    Fig.~\ref{fig:BandStructure}(b) shows that, even in the coplanar case, non-collinear spin order separates up- and down-spin bands.  This marks the distinction between a coplanar non-collinear spin order in a frustrated geometry, like in the present 2D kagome lattice, and a collinear antiferromagnetic order on bipartite lattices: both cases have vanishing net magnetization, but in the latter case the bands remain two-fold spin degenerate, while the former case lifts this Kramers degeneracy.

    Canting the spin ordering gives rise to a remarkably simple band structure whichwhich preserves the large trivial gap of the coplanar case separating the two spin sectors, while adding significant nontrivial gaps. In Fig.~\ref{fig:BandStructure}(c), for $M_s=1\,$eV, these nontrivial gaps are of magnitude $\Delta E\sim 100\ldots 150\,$meV within each spin sector\cite{NoteGaps}, while a large trivial gap separates the two spin sectors and allows for spin-polarized states. Placing the Fermi level in one of the smaller nontrivial gaps, by contrast, may result in topological states such as a Chern insulator. Spin canting therefore induces a spin splitting which neither depends on relativistic effects nor on the distinction between different $d$-orbitals, thus preserving the full orbital degeneracy and a simple band structure.  This makes canted spin textures promising candidates in the quest for unexplored topological phases of matter.
	
% %%%%%%%%%%%%%%%%%%%%%%%%%%%%%%%%%%%%%%%%%%%%%%%%%%
\section{Real and Momentum Space Effective Magnetic Fluxes}
      
    The non-collinear spin order gives rise to an effective magnetic flux felt by an electron\cite{MakhfudzPujolPRL2015} via a gauge field defined along the nearest-neighbor bond $a_{i j}$,
    \begin{equation}   
        \label{PhiTriangle} 
        \Phi_{\Delta,\triangledown}=\sum_{ij\in \Delta,\triangledown}a_{ij}
        \equiv \oint \mathbf{a}\cdot d\mathbf{r}
        =\int d\mathbf{\mathcal{S}}\cdot\left(\nabla\times\mathbf{a}\right)
        \,\text{.}
    \end{equation}
    The sum is over nearest neighbors around an up or down triangle, the $\mathbf{a}$ vector is the continuous form of the gauge field,\cite{MakhfudzPujolPRB2015,AuerbachBook} $d\mathbf{r}$ defines the path along the triangle, and $\mathbf{\mathcal{S}}$ is the surface normal vector. The local spin $\mathbf{S}_{\mathbf{r}} =S(\sin\theta_{\mathbf{r}}\cos\phi_{\mathbf{r}},\sin\theta_{\mathbf{r}}\sin\phi_{\mathbf{r}},\cos\theta_{\mathbf{r}})$ corresponds to the magnetic moment $\mathbf{M}$ in Eq.~(\ref{Hamiltonian}).  A computation for spin-$\frac{1}{2}$ coherent states,\cite{AuerbachBook} appropriate for electrons, gives the gauge field\cite{MakhfudzPujolPRB2015}
    \begin{equation}\label{GaugeField}
        a_{\mathbf{r}\mathbf{r}'}=-\arctan\left[\frac{\sin\left(\phi_{\mathbf{r}'}-\phi_{\mathbf{r}}\right)\sin\frac{\theta_{\mathbf{r}}}{2}\sin\frac{\theta_{\mathbf{r}'}}{2}}{\cos\frac{\theta_{\mathbf{r}}}{2}\cos\frac{\theta_{\mathbf{r}'}}{2}+\cos\left(\phi_{\mathbf{r}'}-\phi_{\mathbf{r}}\right)\sin\frac{\theta_{\mathbf{r}}}{2}\sin\frac{\theta_{\mathbf{r}'}}{2}}\right]
    \end{equation}
    where $\mathbf{r}=(x_i,y_i),\mathbf{r}'=(x_j,y_j)$, denotes the position vector of a lattice site and $a_{ij}\equiv a_{\mathbf{r}\mathbf{r}'}$. For coplanar non-collinear spin order with inter-spin angle 120°, $\theta_{\mathbf{r}}=\theta_{\mathbf{r}'}=\pi/2$ and $\phi_{\mathbf{r}'}-\phi_{\mathbf{r}}=2\pi/3$, the gauge field simplifies to $a_{\mathbf{r}\mathbf{r}'}=-\pi/3$, thus yielding the trivial fluxes $\Phi_{\Delta,\triangledown}=-\pi$ of the time-reversal invariant phase with $\exp(i\pi)=\exp(-i\pi)$.\cite{OhgushiPRB} This result suggests that coplanar non-collinear spin order alone is not sufficient to produce a nontrivial flux that breaks time-reversal symmetry.

    Non-coplanar spin order, by contrast, easily gives nontrivial fluxes, $\Phi_{\Delta,\triangledown}\neq 0, \pm \pi$: still assuming the aforementioned $\phi_{\mathbf{r}'}-\phi_{\mathbf{r}}=2\pi/3$, but with some finite $\theta_{\mathbf{r}}=\theta_0$ for all sites, we obtain $$a_{\mathbf{r}\mathbf{r}'}=-\arctan\left[\frac{\sqrt{3}\tan^2\frac{\theta_0}{2}}{2-\tan^2\frac{\theta_0}{2}}\right]\,\text{.}$$ This gauge field gives a nontrivial effective flux that can be associated with scalar spin chirality\cite{WenWilczekZee, MakhfudzJPCM2018}
    \begin{equation}
        \label{SSchirality} 
        \chi = \mathbf{S}_i\cdot\left(\mathbf{S}_j\times\mathbf{S}_k\right)=-\sfrac{3}{2}\sqrt{3} S^3\cos\theta_0\sin^2\theta_0
        \,\text{,}
    \end{equation}
    where $i, j, k$ are the clockwise labels of three spins on a triangle, and $\theta_i=\theta_0$ on all three sublattices. From the scalar triple product structure, it is clear that the spin chirality vanishes for coplanar non-collinear antiferromagnetic spin order ($\theta_0=\pi/2$), as well as for ferromagnetic order ($\theta_0=0,\pi$). Note that the scalar spin chirality $\chi$ changes sign when the canting is flipped, i.e. the spin texture mirrored by the plane, with $\theta_0\rightarrow \pi-\theta_0$ amounting to the same chirality but with opposite sign. We thus expect two-fold degenerate states with related topological character but opposite chiralities. 

    In the QAHE, the Hall conductivity is quantized in terms of an integer known as the Chern number $C$.  The latter is a topological invariant of the band structure and can be computed from the Berry curvature $\Omega(\mathbf{k})$ in momentum space.\cite{Haldane2004PRL,TKNN} At finite temperature $T$, we have\cite{AHErmp}
    \begin{equation}\label{InterbandChernGeneralT}
	    \Omega(\mathbf{k})=\hbar^2\sum_{n\neq n'}G_{nn'}(\mathbf{k}) \mathrm{Im}\left[\frac{\langle\psi_{n\mathbf{k}}|v_x(\mathbf{k})|\psi_{n'\mathbf{k}}\rangle \langle\psi_{n'\mathbf{k}}|v_y(\mathbf{k})|\psi_{n\mathbf{k}}\rangle}{\left(\varepsilon_{n'\mathbf{k}}-\varepsilon_{n\mathbf{k}}\right)^2}\right]
    \end{equation}
    with $G_{nn'}(\mathbf{k})=f(\varepsilon_{n\mathbf{k}})-f(\varepsilon_{n'\mathbf{k}})$, the Fermi function $f(\varepsilon)=[\exp(\beta(\varepsilon-E_F))+1]^{-1}$, and $\beta={1}/{k_BT}$ the inverse temperature. $\ket{\psi_{n\mathbf{k}}}$ and $\varepsilon_{n\mathbf{k}}$ are the $n^{\mathrm{th}}$ eigenstate and eigenenergy of the $\mathbf{k}$-space Hamiltonian, and $\mathbf{v}(\mathbf{k})=\frac{1}{\hbar}\nabla_{\mathbf{k}} H(\mathbf{k})$ is the velocity operator. The Berry curvature can be interpreted as resulting from a non-Abelian Berry monopole in $\mathbf{k}$-space. 
	
    \begin{figure*}
        \includegraphics[width=7in]{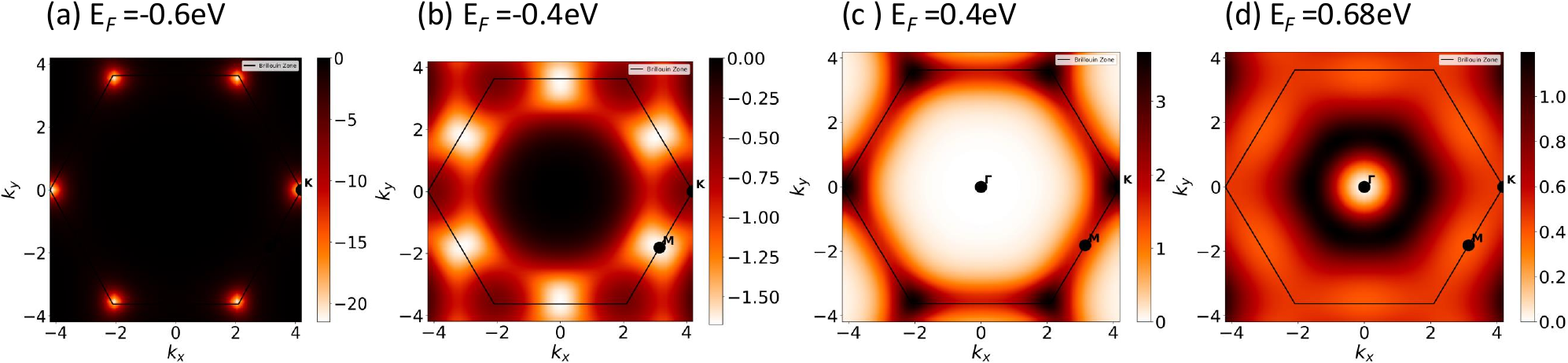}
	    \caption{Heat maps of the dimensionless Berry curvature $\Omega(k_x, k_y)$ for the parameter set of Fig.~\ref{fig:BandStructure}(c). The Fermi energy lies within the nontrivial gaps of the band structure, and is located in the spin-up sector for (a) and (b), and in the spin-down sector for (c) and (d).}
        \label{fig:BerryCurvatureField}
    \end{figure*}
	
    The Hall conductivity $\sigma_{xy}$ is related to the integral of Berry curvature over the Brillouin zone
    \begin{equation}
       \label{InterbandChernT=0}
        \sigma_{xy}=\frac{e^2}{h}\,\frac{1}{2\pi}\int\limits_{\mathrm{BZ}} d^2{k}\;\Omega(\mathbf{k})=\frac{e^2}{h}\,\frac{\Phi_B}{2\pi}
        \,\text{.}
    \end{equation}
    The quantization of the Hall conductivity is rooted in the Berry curvature in reciprocal space: its integral over the Brillouin zone is an integer -- the Chern number $C$ -- making the Hall conductivity an integer multiple of the conductance quantum, $\sigma_{xy}=\frac{e^2}{h}C$.\cite{Haldane2004PRL}

% %%%%%%%%%%%%%%%%%%%%%%%%%%%%%%%%%%%%%%%%%%%%%%%%%%
%\section{QAHE from Canted Antiferromagnetic Order in $d$-Electron System: Spin- Momentum Locking}
\section{Canted Antiferromagnetic Order and Spin-Momentum Locking}

    \begin{figure}
        \includegraphics[width=3.2in]{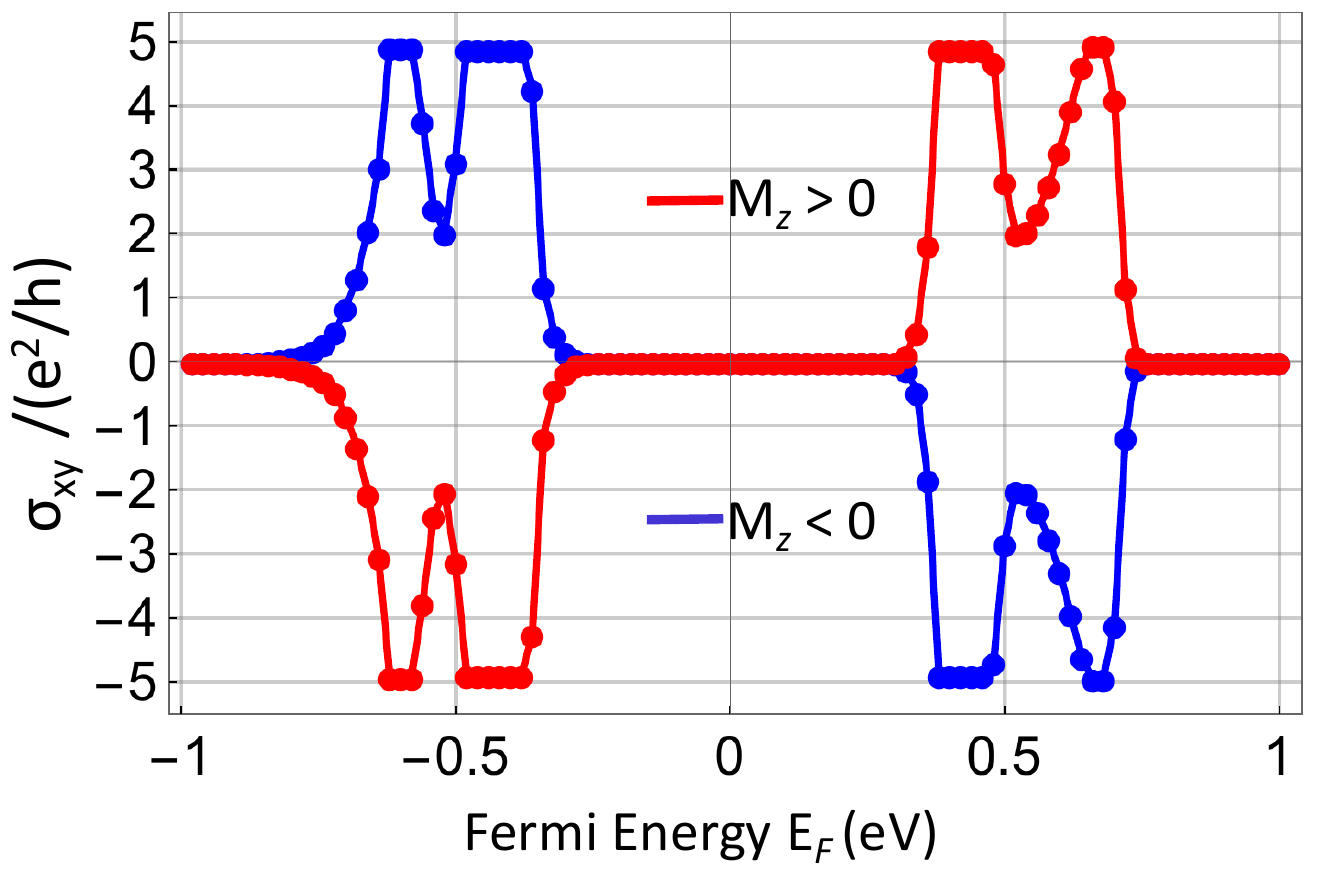}
        \caption{Dimensionless Hall conductivity $\sigma_{xy}/(e^2/h)$ as a function of the Fermi energy $E_F$, at $T=0.1\, \mathrm{meV}=1.16\,\mathrm{K}$ and for 120$^\circ$ canted (non-coplanar) antiferromagnetic order. Same parameters as in Fig.~\ref{fig:BandStructure}(c), with $M_z=0.6\,\mathrm{eV}$ (red) and $M_z=-0.6\,\mathrm{eV}$ (blue), to illustrate symmetry. For some values of $E_F$, plateaus occur at the maximal Chern number $C=\pm 5$. Dots correspond to numerical data, solid lines are guides to the eye, and arrows indicate the spin sectors.}
        \label{fig:ChernNumber}
    \end{figure}
   
    We first examine the Berry curvature for the particular parameter set of Fig.~\ref{fig:BandStructure}(c), which leads to the most remarkable Hall conductance. 
    Fig.~\ref{fig:BerryCurvatureField} displays the Berry curvature $\Omega(\mathbf{k})$ over the Brillouin zone.  The peaks in this landscape are governed by the Fermi energy, and a direct comparison to Fig.~\ref{fig:BandStructure}(c) shows that the regions in which they occur correspond to the locations where the nontrivial gaps are tightest and where the curvature of the band is largest.

    The exact Hall conductivity $\sigma_{xy}$ (in units of the conductance quantum $e^2/h$) is computed numerically from the full $30 \times 30$ Hamiltonian matrix via Eq.~(\ref{InterbandChernGeneralT}). 
    Fig.~\ref{fig:ChernNumber} shows that quantization in terms of an integer multiple of $e^2/h$ arises only if the Fermi energy lies within a gap, while for Fermi energies within a band, $\sigma_{xy}$ is not quantized. The integer multiple is called the Chern number $C$. However, for non-collinear coplanar spin order with isotropic Slater-Koster integrals, $C$ vanishes everywhere, and this behavior persists in the anisotropic case since the net magnetization still vanishes.

    In our model (\ref{Hamiltonian}), a nonzero $C$ occurs only for non-coplanar non-collinear spin order, since only canted spin order allows for nonzero scalar spin chirality. When all onsite energies are degenerate, a relatively large $C=\pm 5$ plateau is obtained.  
    It lies five times higher than the Chern plateaus found in ferromagnetic kagome systems,\cite{SCZhangPRL,Nature2018} with all five $d$-orbitals contributing equally to the total Chern number.  
    As pointed out before, this requires nontrivial real-space fluxes $\Phi_{\Delta,\triangledown}\neq 0, \pm \pi$, such as provided by canted ordering, but absent in the coplanar non-collinear antiferromagnetic as well as in the (collinear) ferromagnetic case which therefore both result in trivial fluxes. This result offers, at least in principle, an alternative avenue towards a large-Chern QAHE in a kagome system, occurring naturally, i.e. without relativistic effects\cite{ZhangPRL} or heterostructure engineering.\cite{BernevigPRL}

    The two spin sectors in Fig.~\ref{fig:ChernNumber} carry opposite Chern numbers, reflecting the fact that the scalar spin chirality conserves its magnitude but changes sign for polar angles $\theta_0$ and $\pi-\theta_0$.
    By tuning the Fermi energy or the filling factor,  one of the two counter-propagating chiral edge states of a given spin state can be switched on or off, thus producing half of the edge states of the quantum spin Hall effect \cite{KaneMelePRL} in two dimensions. The two spin sectors always contribute with opposite Chern numbers, since the underlying chiral edge states propagate in opposite directions: spin and momentum are thus locked, such that the two spin sectors never carry a Chern number of the same sign. Due to this spin-momentum locking, $|C| = 5$ is the upper limit for the Chern number in our model.

    This value turns out to be robust with respect to varying the canting angle. As shown in Ref.~\onlinecite{OhgushiPRB} for $s$-electrons on a kagome lattice, the Chern number only depends on the sign of the flux in the triangle, or more specifically on that of its sine, $C=\mathrm{sign}\left[\sin(\Phi)\right]$.   Therefore, the precise value of the flux for a given canting angle $\theta$ is unimportant -- as long as the angles zero and $\pi$ are avoided -- such that each orbital contributes with $C=\pm 1$ to the overall Chern number.

    The Hall conductivity presented in Fig.~\ref{fig:ChernNumber} was obtained for a temperature of $T=0.1\, \mathrm{meV}=1.16\,\mathrm{K}$. Even though strictly speaking, the integer quantization of the Chern number requires zero temperature, supplementary calculations show that, in fact, the quantization persists to good accuracy up to temperatures of several tens of Kelvin.  For even higher temperatures, in the range of $100-250$ K, the conductance still retains the displayed two-peak structure, but the plateaus are found to be significantly rounded.
    
% %%%%%%%%%%%%%%%%%%%%%%%%%%%%%%%%%%%%%%%%%%%%%%%%%%
\section{Low-Energy Theory and Topological Transition}

    To gain conceptual insight beyond numerics, and in particular to understand the qualitative features of Fig.~\ref{fig:ChernNumber}, we derive a low-energy effective theory for the $d$-electrons. Around the Dirac point which, in the absence of spin order, is located at $K$ in the first Brillouin zone, the low-energy theory contains two parts: (i) the Dirac Hamiltonian, and (ii) a Zeeman term.
    
    \subsection{Dirac Hamiltonian}
    
     \begin{figure}
		\includegraphics[angle=0,origin=c, scale=0.23]{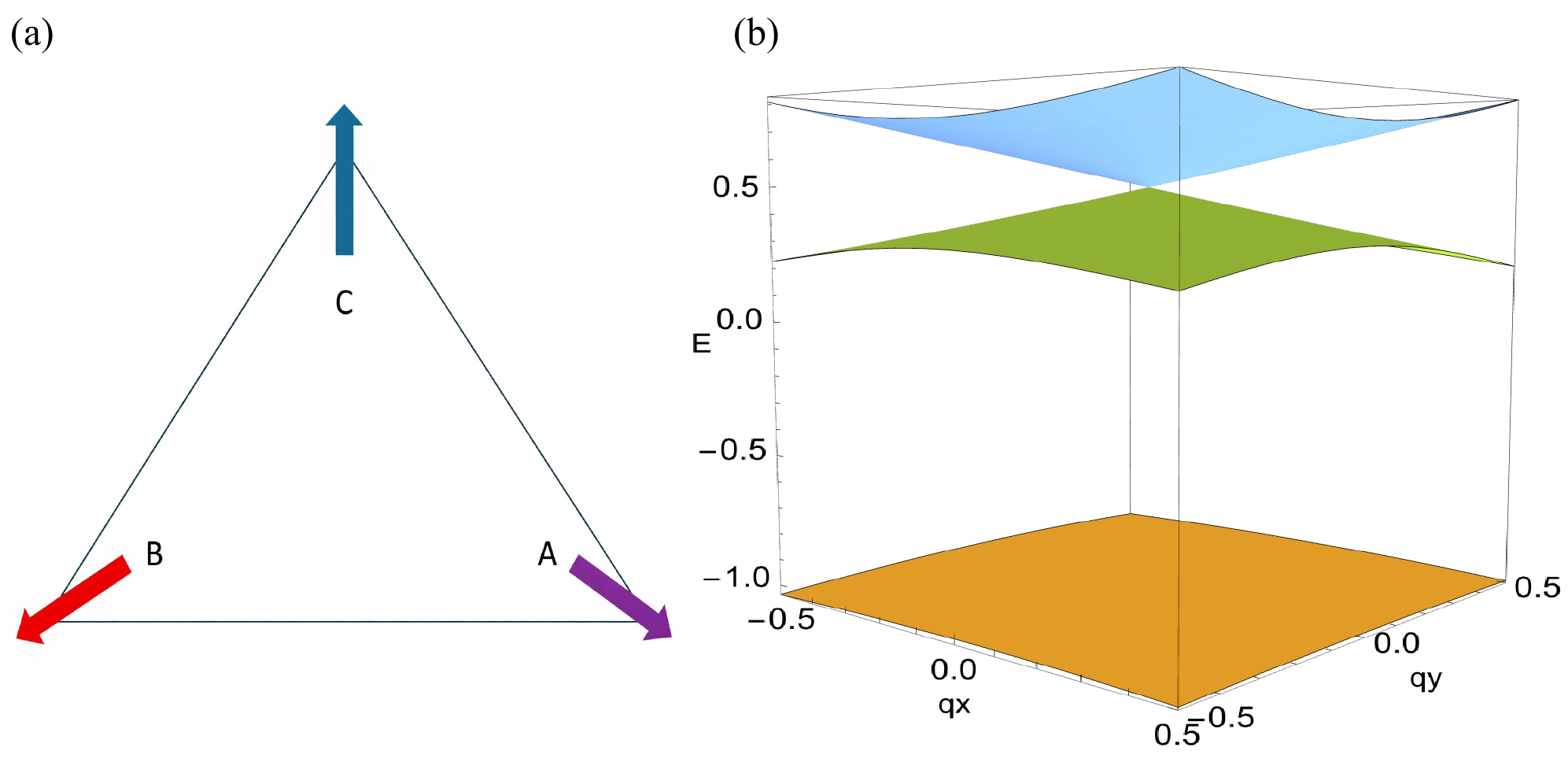}
		\caption{(a) The unit cell of a 2D kagome lattice with its three sublattices and the corresponding spin vectors.
        (b) Band structure of the low-energy Hamiltonian $h(\mathbf{K})+h_{\mathrm{Dirac}}(\mathbf{q})$ with its Dirac point.}
        \label{mercedes/Dirac}
	\end{figure}
    In the isotropic limit, we have $t=V_{dd\sigma}=V_{dd\delta}=V_{dd\pi}$. The kinetic part of the Hamiltonian in Fourier space then takes the form of a $3\times 3$ matrix in sublattice space,
    \begin{equation}
        h(\mathbf{k})=t    
        \begin{pmatrix}
        0 & \cos\mathbf{k}\cdot\mathbf{d}_{AB} & \cos\mathbf{k}\cdot\mathbf{d}_{CA}\\
        \cos\mathbf{k}\cdot\mathbf{d}_{AB} & 0 & \cos\mathbf{k}\cdot\mathbf{d}_{BC}\\
        \cos\mathbf{k}\cdot\mathbf{d}_{CA} & \cos\mathbf{k}\cdot\mathbf{d}_{BC} & 0
        \end{pmatrix}
        .
    \end{equation}
    The unit vectors $\mathbf{d}_{AB},\mathbf{d}_{BC},\mathbf{d}_{CA}$ between nearest-neighbor sites (or sublattices) are illustrated in Fig.~\ref{mercedes/Dirac}. An expansion around the Dirac point $\mathbf{k}=\mathbf{K}+\mathbf{q}$ yields
    \begin{equation}
       h(\mathbf{k})=h(\mathbf{K})+h(\mathbf{q})
    \end{equation}
    where $h(\mathbf{K})$ is the Hamiltonian at the Dirac point $\mathbf{K}=(\frac{4\pi}{3}, 0)$. In addition to the triangle in the unit cell and the three sublattices, Fig.~\ref{mercedes/Dirac} also shows their spin orientation projected onto the lattice plane. Expanding to linear order in $\mathbf{q}$, i.e. using
    \begin{align*}
        \cos[(\mathbf{K}+\mathbf{q})\cdot\mathbf{d}]
        &=\cos\mathbf{K}\cdot\mathbf{d}\,\cos\mathbf{q}\cdot\mathbf{d}\,-\sin\mathbf{K}\cdot\mathbf{d}\,\sin\mathbf{q}\cdot\mathbf{d}
        \\
        &\simeq \cos\mathbf{K}\cdot\mathbf{d}\,-\mathbf{q}\cdot\mathbf{d}\,\sin\mathbf{K}\cdot\mathbf{d}
        ,
    \end{align*}
    (for $\mathbf{d}=\mathbf{d}_{AB}$, $\mathbf{d}_{BC}$, or $\mathbf{d}_{CA}$), yields a constant part of
    \begin{equation}
        h(\mathbf{K})=\frac{t}{2}
        \begin{pmatrix}
            \begin{array}{rrr}
                0 & -1 & +1\\
                -1 &  0 & +1\\
                +1 & +1 & 0
            \end{array}
        \end{pmatrix}
        .
    \end{equation}
    With the shorthand $s_{ij}=\sin\mathbf{K}\cdot\mathbf{d}_{ij}$,
    the linear term reads\footnote{Here, we use nearest-neighbor vectors of length $1/2$ instead of $1$ as in Ref.~\onlinecite{MakhfudzPRB2024}.}
    \begin{equation}
        h(\mathbf{q})=t    
        \begin{pmatrix}
            0 & -\mathbf{q}\cdot\mathbf{d}_{AB}s_{AB} & -\mathbf{q}\cdot\mathbf{d}_{CA}s_{CA}\\
            -\mathbf{q}\cdot\mathbf{d}_{AB}s_{AB} & 0 & -\mathbf{q}\cdot\mathbf{d}_{BC}s_{BC}\\
            -\mathbf{q}\cdot\mathbf{d}_{CA}s_{CA} & -\mathbf{q}\cdot\mathbf{d}_{BC}s_{BC} & 0
        \end{pmatrix}
        .
    \end{equation}
    Introducing the matrices
    \begin{equation}
        J_x=\begin{pmatrix}
                \begin{array}{rrr}
                    0 & -2 & -1\\
                    -2 & 0 & -1\\
                    -1  &  -1  & 0
                \end{array}
            \end{pmatrix}, 
        \quad
        J_y=\begin{pmatrix}
                \begin{array}{rrr}
                    0 & 0 & 1\\
                    0 & 0 & -1\\
                    1  &  -1  & 0
                \end{array}
            \end{pmatrix},
    \end{equation}
    allows us to reduce the linear correction to
    \begin{equation}
        h(\mathbf{q})=\frac{\sqrt{3}}{8}t\left(q_xJ_x+\sqrt{3}q_yJ_y\right).
    \end{equation}
    
    The band structure around the Dirac point is shown on the right side of Fig.~\ref{mercedes/Dirac}. It features three bands, one of which is dispersionless (at the bottom), while the other two touch each other at the Dirac point. 
    
    \subsection{Zeeman term}
    
    We now consider the Zeeman term, which is off-diagonal in spin space, but diagonal in sublattice space.  
    With $\mathbf{s}=\frac{\hbar}{2}\mathbf{\sigma}$ and $\mathbf{\sigma}=(\sigma_x,\sigma_y,\sigma_z)$ denoting the vector of Pauli spin matrices, we have
    \begin{equation}
        h_{\mathrm{Zeeman}}=-\sum_{i\in A,B,C} \mathbf{M}_i\cdot\mathbf{s}\,.
    \end{equation}
    For the spin orientation of Fig.~\ref{mercedes/Dirac}, and including the uniform out-of-plane contribution $M_z$, this yields
    \begin{equation}
        h_{\mathrm{Zeeman}}=-\frac{\hbar}{2} 
        \left(J_0\, M_z \sigma_z
        +\sum_{i\in A,B,C}J_i\,\mathbf{M}^i_{\perp}\cdot\sigma_{\perp}   %\left(\mathbf{M}^A_{\perp}J_A+\mathbf{M}^B_{\perp}J_B+\mathbf{M}^C_{\perp}J_C\right)
        \right).
    \end{equation}
    $\sigma_{\perp}=(\sigma_x,\sigma_y,0)$ and $\mathbf{M}^A_{\perp}=(M^A_x,M^A_y,0)=M_{\perp}(\frac{\sqrt{3}}{2},-\frac{1}{2},0)$ (and analogously for sublattices $B$ and $C$), denote the in-plane components of $\mathbf{M}$ and $\mathbf{\sigma}$, while a non-vanishing $M_z$ is responsible for the canting of the spin texture.

    The $J$-matrices are identity matrices in sublattice space, with $J_i$ selecting sublattice $i$ and $J_0$ any of the three sublattices, i.e. $\left(J_0\right)_{ij}=\delta_{ij}$, $\left(J_{A}\right)_{ij}=\delta_{ij}\delta_{iA}$, or equivalently
    \begin{equation}
        \begin{aligned}
        J_0 &=
        \begin{pmatrix}
        1&0&0\\ 0&1&0\\ 0&0&1
        \end{pmatrix}
        &
        J_A &=
        \begin{pmatrix}
        1&0&0\\0&0&0\\0&0&0
        \end{pmatrix}
        \\[10pt]
        J_B &=
        \begin{pmatrix}
        0&0&0\\0&1&0\\0&0&0
        \end{pmatrix}
        &
        J_C &=
        \begin{pmatrix}
        0&0&0\\0&0&0\\0&0&1
        \end{pmatrix}
        .
        \end{aligned}
    \end{equation}

    With $\sigma_0$ denoting the identity in spin space, we can now define the $6\times 6$ massive Dirac Hamiltonian
    \begin{align}
        h_{\mathrm{MD}}(\mathbf{q})
        &=h_{\mathrm{Dirac}}(\mathbf{q})+h_{\mathrm{Zeeman}} 
        \,,
        \\
        h_{\mathrm{Dirac}}(\mathbf{q})&=\left[h(\mathbf{K})+h(\mathbf{q})\right]\,\sigma_0
        \,.
    \end{align}

    Defining a vector of $3\times 3$ matrices in sublattice space,
    \begin{align}
        \mathbf{M}_J &= (M^x_J, M^y_J, M^z_J) 
        \\
        &= \left(
            (J_A - J_B)\frac{\sqrt{3}}{2} M_{\perp},
            \; \left(J_C - \frac{J_A + J_B}{2}\right) M_{\perp},
            \; J_0 M_z
        \right),
        \nonumber  
    \end{align}
    we may finally express the massive Dirac Hamiltonian as 
    \begin{equation}
        \label{DiracHamiltonianS}
        h_{\mathrm{MD}}(\mathbf q)
        = \left[ h(\mathbf K)
        + \frac{\sqrt{3}}{8} t
        \left(q_x J_x + \sqrt{3} q_y J_y\right)\right]\sigma_0
        - \frac{\hbar}{2}\,\mathbf{M}_J\cdot\mathbf{\sigma}
        ,
    \end{equation}
    with the mass in the last term originating from the Zeeman contribution.

\subsection{Diagonalization of the Hamiltonian}

    A unitary transformation diagonalizes the Hamiltonian in sublattice space, 
    \begin{equation}
        h_{\mathrm{MD}}(\mathbf{q})\;\longrightarrow\; h^{\mathrm{diag}}_{\mathrm{MD}}(\mathbf{q})
        =U^{\dagger}h_{\mathrm{Dirac}}(\mathbf{q})U \,+U^{\dagger}h_{\mathrm{Zeeman}}U
        \,.
    \end{equation}
    Note that $h_{\mathrm{Zeeman}}$ was already diagonal in sublattice space and is thus invariant. We obtain
    \begin{align}
         h^{\mathrm{diag}}_{\mathrm{MD}}(\mathbf q)
        &=\mathrm{diag}\!\left(H_1(\mathbf q),H_2(\mathbf q),H_3(\mathbf q)\vphantom{\frac{1}{1}}\right)
        \\
        H_i(\mathbf q)&=E_{ii}(\mathbf q)+
        \, \mathbf d^{\,i}(\mathbf q)\!\cdot\!\sigma 
    \end{align}
    where $\mathbf{d}^{{i}}(\mathbf{q})$ are the Bloch vectors for each state.

 \subsection{Topological Transition}
 
    After normalization, the vector $\hat{\mathbf{d}}(\mathbf{q})=\mathbf{d}(\mathbf{q})/|\mathbf{d}(\mathbf{q})|$ lives on a Bloch sphere in $\mathbf{k}$-space. The Hall conductivity $\sigma_{xy}=\frac{e^2}{2\pi\hbar}\,C$ in terms of the Chern number can then be computed from an integral over the first Brillouin zone involving the partial derivatives of the Bloch vectors in $\mathbf{q}$-space.\cite{XLQiPRB}
    \begin{align}
        \label{HallconductivityIntegral}
        %\sigma_{xy}&=\frac{e^2}{2\pi\hbar}C
        %\\
        \sigma_{xy}& =-\frac{1}{8\pi^2} \frac{e^2}{\hbar} \int\limits_{\mathrm{1BZ}}d^2 q \;\hat{\mathbf{d}}(\mathbf{q}) 
        \cdot\left[
        \partial_{q_x}\hat{\mathbf{d}}(\mathbf{q}) \times
        \partial_{q_y}\hat{\mathbf{d}}(\mathbf{q}) \right] 
        \,.
    \end{align}
 
    In principle, the Bloch vectors $\mathbf{d}^{{i}}(\mathbf{q})$ can be analytically obtained from Eq.~(\ref{DiracHamiltonianS}); the resulting expression is, however, rather complicated since it involves the solution of a cubic equation. Note that the integral in Eq.~(\ref{HallconductivityIntegral}) resembles the Skyrmion number, which simply counts how many times the unit vector $\hat{\mathbf{d}}$ winds around the Bloch sphere in $\mathbf{k}$ space as one sweeps the Brillouin zone. Similarly, the Chern number can be interpreted as the solid angle (in units of $4\pi$) covered by $\hat{\mathbf{d}}$ when scanning through the Brillouin zone. 

    Another important observation is that the $z$-components of all three $\mathbf{d}^{{i}}(\mathbf{q})$ are identical and $\mathbf{q}$-independent, 
    \begin{equation}
        \label{dz}
        d^i_z(\mathbf{q}) = -\frac{\hbar M_z}{2}
        \,.
    \end{equation}
    The Chern number in Eq.~(\ref{HallconductivityIntegral}) can thus be written as 
    %{\color{red} ((check if it is $1/4\pi$ not $1/4\pi^2$ ---> ok))}
    \begin{equation}
        C=-\frac{1}{4\pi} \int\limits_{\mathrm{1BZ}}d^2{q}
        \;\hat{d}^{{i}}_z(\mathbf{q})
        \left[\frac{\partial\hat{d}^{{i}}_x(\mathbf{q})}{\partial q_x}
        \frac{\partial\hat{d}^{{i}}_y(\mathbf{q})}{\partial q_y}  -\frac{\partial\hat{d}^{{i}}_y(\mathbf{q})}{\partial q_x}   
        \frac{\partial\hat{d}^{{i}}_x(\mathbf{q})}{\partial q_y}\right]
        \,\text{.}
        \label{chernintegral}
    \end{equation}
    Due to the equivalence of the three sublattices, the Chern number may be calculated from Eq.~(\ref{chernintegral}) for any value of $i$.
    More importantly, it changes sign with $d^i_z$ which, according to Eq.~(\ref{dz}), does not depend on $\mathbf{q}$ but solely on $M_z$.  This demonstrates that the Chern number changes sign with the out-of-plane component $M_z$.

    For $\theta\simeq \frac{\pi}{2}$, i.e. close to the equator, a simple expression for the real-space spin $s=\frac{1}{2}$ Berry phase can be obtained from a continuum formula,\cite{AuerbachBook} 
    \begin{equation}
        \Phi = s\oint d\tau \dot{\phi}(1-\cos\theta)=\pi(1-\cos\theta)
         ,
    \end{equation}
    which asymptotically reproduces $\Phi_{\Delta,\triangledown}$ of Eqs.~(\ref{PhiTriangle}) and (\ref{GaugeField}) in the limit of $|\tan\theta|\to\infty$. According to the result of the flux model,\cite{OhgushiPRB} for Fermi energies within a nontrivial gap, the Chern number for the kagome lattice with degenerate onsite energies for all five $d$-orbitals is given by
    \begin{equation}
        \label{C5}
	    C=\pm 5\cdot\mathrm{sign}\left[\sin(\pi\left(1-\cos\theta\right))\right]
        \,\text{,}
    \end{equation}
    where $\pm$ applies to the bands in the upper (lower) sector. Eq.~(\ref{C5}) is an upper bound for the maximally achievable Chern number in $d$-electron systems with canted spin order.  Higher Chern numbers are ruled out because of spin-momentum locking, which impedes both spin sectors from contributing with the same sign to the overall Chern number.

% %%%%%%%%%%%%%%%%%%%%%%%%%%%%%%%%%%%%%%%%%%%%%%%%%%
\section{Candidate Materials for Experiments}
        
    The discovery of metal-organic frameworks (MOFs) by Kitagawa, Robson and Yaghi\cite{YaghiNature} was honored by last year’s Nobel Prize in Chemistry. 
    While the original MOFs were mainly three-dimensional porous architectures, our present article is motivated by a more recent development within the MOF family: the materials we consider are two-dimensional kagome lattices hosting transition metal ions,\cite{Shaiek} resulting in a chemical and electronic structure more complex than that of graphene, in which the $d$-orbitals of the transition metals play a key role. 
    From the perspective of the tight-binding model, the key parameters are the Slater-Koster integrals, since they strongly determine the form of the band structure. The second important condition is the feasibility of the canted antiferromagnetic order. In our model (\ref{Hamiltonian}), the magnetic order is implicitly assumed to arise from an appropriate combination of (presumably frustrated) spin interactions that stabilize such non-collinear non-coplanar spin configurations. This is sufficient but not necessary; one may also assume placing the $d$-electron kagome monolayer on top or in the vicinity of an appropriate substrate that stabilizes the desired canted spin order.
    
    Various compounds are known to encompass $d$-electron kagome systems with interesting magnetic and electronic structure properties.\cite{CominNatureMaterials0, GiefersNicol,SinghNPJqm,Nature,KimLiuPRB2023FlatBandsIntercalation} 
    In general, they can be divided into two classes: inorganic kagome metals and MOFs. Characteristically, inorganic kagome metals have large bandwidths and small onsite energy splitting, while MOFs usually have bandwidths smaller than the onsite energy splitting. When aiming for applications to real materials, with parameters for a representative kagome metal compound, our theory underlines the role of the topological properties of the electron bands, while onsite energies play only a secondary role in shifting the bands. More material specific investigations for different kagome MOFs with varying onsite energies will be reserved for a future study.  
    In general, the Slater-Koster integrals are anisotropic and obey an empirical rule fulfilled by a large class of materials:\cite{NavalPRB} $|V_{\sigma}|>|V_{\pi}|>|V_{\delta}|$, with alternating sign,\cite{CominNatureMaterials0} i.e. $\mathrm{sign}(V_{dd\sigma})=-\mathrm{sign}(V_{dd\pi})= \mathrm{sign}(V_{dd\delta})$.

    \begin{figure}
        \includegraphics[width=\columnwidth]{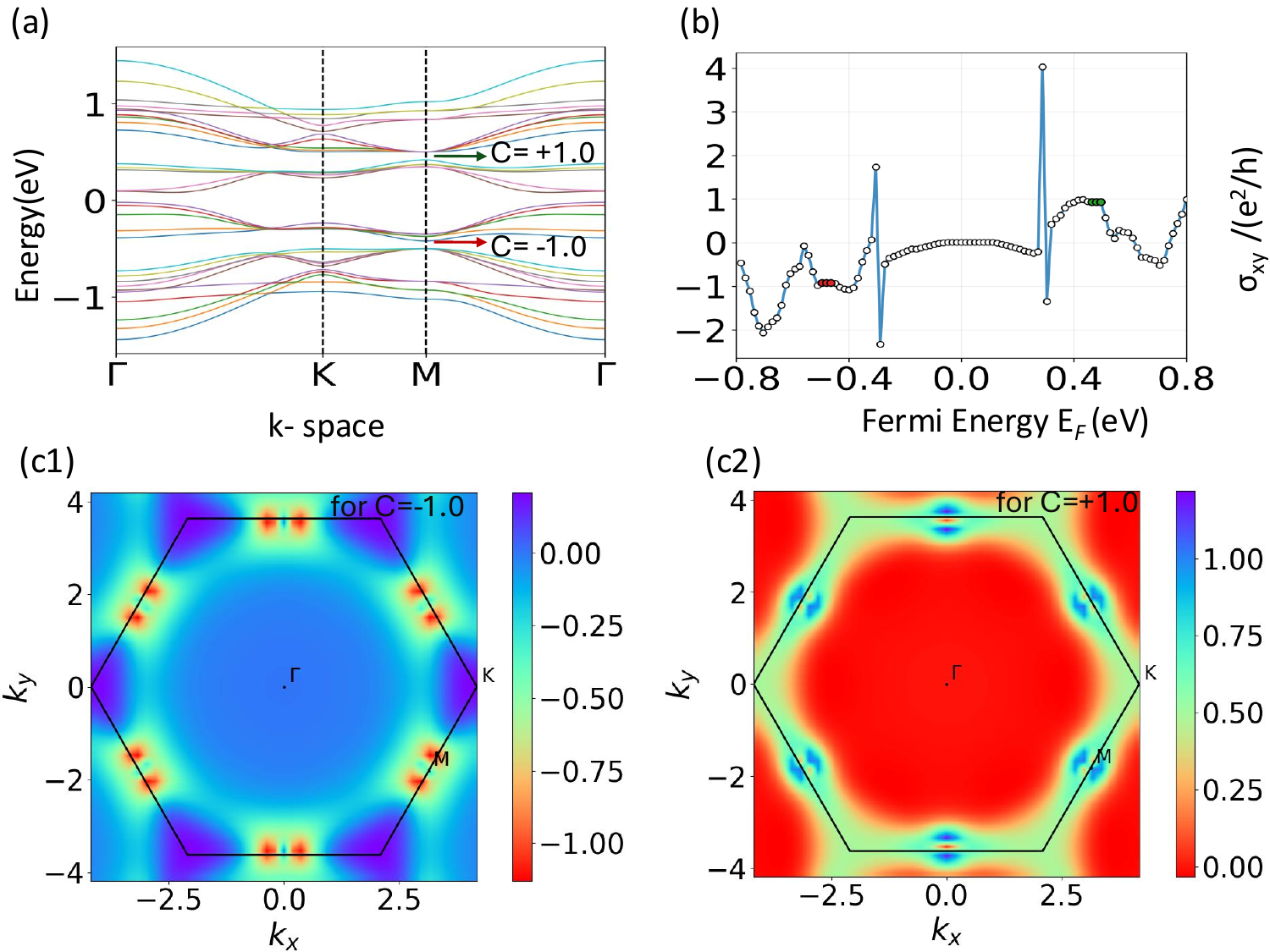}
	    \caption{
	    Predictions for a single kagome layer within FeSn, with realistic Slater-Koster integrals $V_{dd\pi}=0.50$, $V_{dd\delta}=-0.40$, $V_{dd\sigma}=-0.65$ (in eV).  All other parameters as in Figs.~\ref{fig:BandStructure}(c), \ref{fig:BerryCurvatureField}, and \ref{fig:ChernNumber}.  We assume a hypothetical canted magnetic ordering characterized by $M_s=1$, $M_z=0.6$ (in eV).
        Two true nontrivial gaps with Chern numbers $C = \pm1$ are found in (a)-(b).   Additional peaks arise due to nontrivial gaps masked by other bands.}
        \label{fig:RealMaterialsResult}
    \end{figure}
    
    In iron–tin (FeSn), the $d$-electrons reside on Fe atoms, forming stacked two-dimensional kagome layers, with ferromagnetic ordering within each layer, and antiferromagnetic ordering between layers. We use a realistic parameter set which is thought to be representative for a broad class of kagome $d$-electron systems, with Slater-Koster integrals corresponding to  {\sl ab initio} calculations\cite{CominNatureMaterials0} for FeSn, and $M_s=1$\,eV matching the observed splitting between $\uparrow$ and $\downarrow$-spin bands. 
    However, instead of the actual ferromagnetic order within the layer, we assume a hypothetical out-of-plane tilt of the magnetic moments by introducing a canted $M_z=0.6$\,eV.
    
    Figure~\ref{fig:RealMaterialsResult} shows the band structure, Berry curvature, and Chern number obtained for such a FeSn kagome monolayer with canted magnetic ordering. 
    In comparison to the idealized case, presented in Fig.~\ref{fig:BandStructure}, the band structure is far more complicated, as a result of the lifted orbital degeneracy. It nonetheless still hosts nontrivial gaps of about 50 meV, which are smaller than in the isotropic case for the same canting angle. Away from the ferromagnetic ($\theta=0^{\circ}$) and coplanar ($\theta=90^{\circ}$) limits, a significant gap can still be obtained by varying $\theta$, while keeping all other parameters fixed.
    The maximum of about 120 meV occurs for $\theta=35\degree$, and increases with the Zeeman exchange field $M$. In real materials, the latter may be larger than the 1.0 eV used in our calculation. 
    
    The Berry curvature in Fig.~\ref{fig:RealMaterialsResult} remains six-fold symmetric, but displays a more nuanced structure than in the idealized case of Fig.~\ref{fig:BerryCurvatureField}: here, the sign alternates in at least one of the two sectors of the Brillouin zone, while it remains constant in the idealized case. More importantly, the observed Chern numbers drop from 5 to $|C|=1$ for free nontrivial gaps, i.e. gaps not masked by any bands. The conductivities still contain contributions from all $d$-orbitals, but their individual Chern numbers alternate in sign, as a consequence of the alternating Slater-Koster integrals.  For an odd number of orbitals, the cancellation is incomplete, resulting in $C=1$.

    For nontrivial gaps masked by other bands, additional peaks in Hall conductivity may arise, resulting in near-integer values.  In Fig.~\ref{fig:RealMaterialsResult}(b), this is visible e.g. near $C=-2$ at around $E_F=-0.7$eV.  Such features imply the presence of chiral edge states accompanied by some (transverse) bulk conduction by the masking bands. Note that the Chern number flips sign as $M_z$ is reversed, even for parameter sets corresponding to real materials, thus preserving the qualities of the isotropic case. 

% %%%%%%%%%%%%%%%%%%%%%%%%%%%%%%%%%%%%%%%%%%%%%%%%%%
\section{Nontrivial Gap as a function of the Canting Angle in Real Materials}

    As already mentioned, real $d$-electron materials are characterized by anisotropic Slater-Koster integrals with alternating signs. This gives rise to a relatively complicated band structure where bands tend to cross each other. A numerical analysis of the gap as a function of the canting angle and the Slater–Koster integrals would therefore be valuable.

\begin{figure}
        \includegraphics[width=\columnwidth]{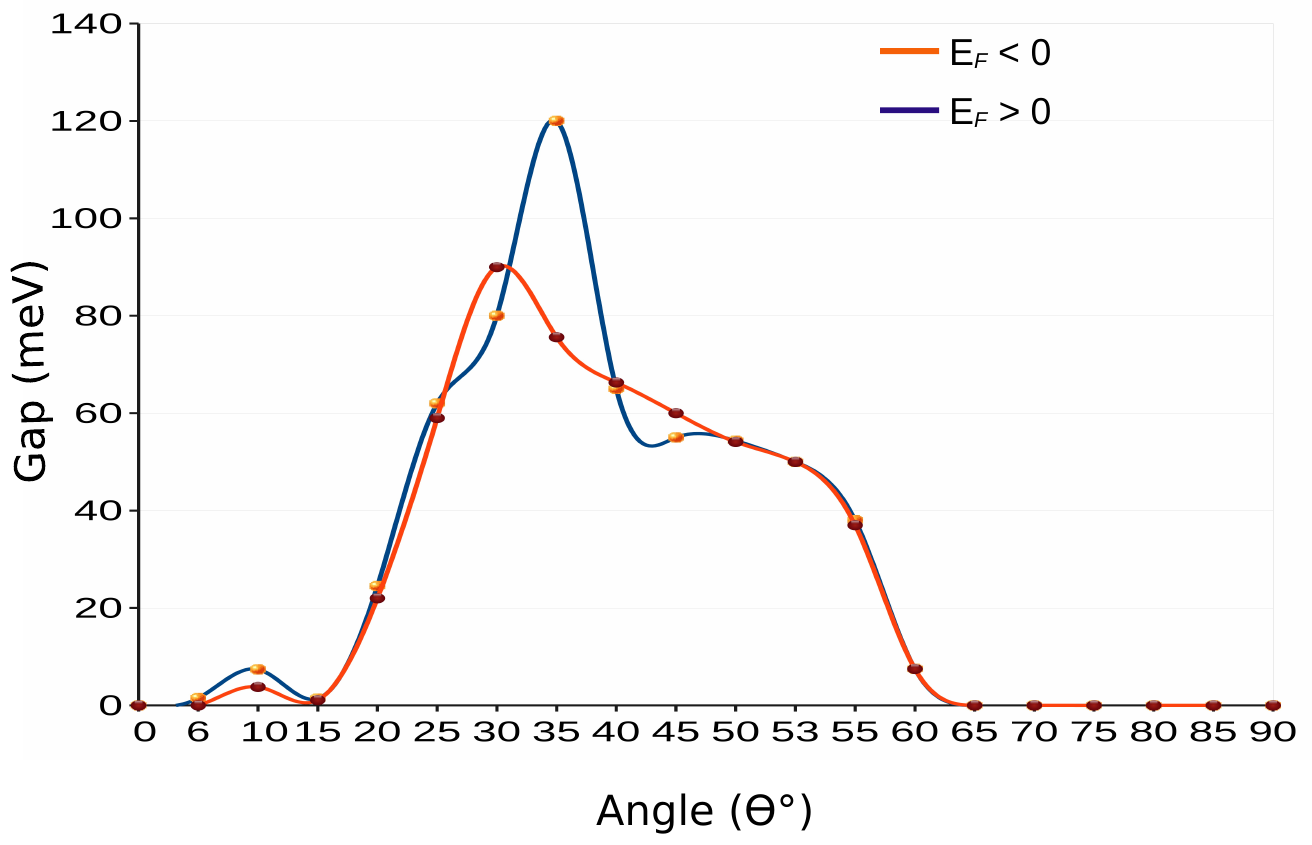}
		\caption{Variation of the nontrivial gap $\Delta_{nt}$ with the canting angle $\theta$ (in degrees) for the parameter set of Fig.~\ref{fig:RealMaterialsResult}, corresponding to FeSn.}
        \label{fig:GapVsCantingAngle}
    \end{figure}

    The numerically computed nontrivial gap $\Delta_{nt}$ as a function of the canting angle, for angles $0^{\circ}\leq\theta\leq 90^{\circ}$ is shown in Fig.~\ref{fig:GapVsCantingAngle}. We observe that the gap neither opens at $0^{\circ}$ nor closes (or switches sign) at $90^{\circ}$, but instead appears only over a finite range of canting angles. Its dependence on the canting angle is non-monotonic and generally involves a maximum somewhere between the north pole and the equator of the Bloch sphere. 
    The maximal gap can be substantial: in the present example, a Zeeman exchange field of $M = 1.0\,\mathrm{eV}$ yields a maximum gap exceeding $80\,\mathrm{meV}$, which is large compared to the nontrivial gap induced by spin--orbit coupling.
    
    Indeed, as demonstrated in our previous work \cite{MakhfudzPRB2024}, a spin-orbit coupling of $200\,\mathrm{meV}$ -- which for real materials is unusually large -- only induces a nontrivial gap of about $20\,\mathrm{meV}$.  On the contrary, typical Zeeman exchange fields can reach $3\,\mathrm{eV}$ in real materials, which would entail even larger nontrivial gaps for a spin order canted to some finite angle. 
    
    Our numerical calculations confirm that the nontrivial gap increases with the Zeeman exchange field $M$, highlighting the advantage of canted spin order for the generation of a QAHE on larger energy scales.
    
% %%%%%%%%%%%%%%%%%%%%%%%%%%%%%%%%%%%%%%%%%%%%%%%%%%
\section{Discussion and Conclusions}
    
    We have developed a comprehensive tight-binding model for $d$-electrons on a kagome lattice.  The model treats all $d$-orbitals fully and on equal footing, making it applicable to a wide range of materials, including both inorganic and organic metals.
    We used our model to investigate the quantum anomalous Hall effect induced in the system of $d$-electrons by a canted spin order on the underlying kagome lattice with antiferromagnetic couplings.  This mechanism is fully operational without spin-orbit coupling. In the idealized isotropic limit, we observe a quantum anomalous Hall effect of maximal Chern number. By contrast, realistic parameters for existing materials exhibit highly anisotropic hopping integrals that depend on the orbital type and often alternate in sign, leading to smaller Chern numbers.  Our findings show that canted spin order not only increases the energy range in which the quantum anomalous Hall effect occurs, but also opens the possibility for a quantum anomalous Hall effect with larger Chern numbers than the usual $|C|=1$.  This may pave the way to dissipationless quantum transport in such multi-orbital materials.
   
% %%%%%%%%%%%%%%%%%%%%%%%%%%%%%%%%%%%%%%%%%%%%%%%%%%
\section{Acknowledgments}

    I.M.~thanks M.~Diouf, N.~Saidi, R.~Saidi, M.~Ngom, C.~Ngbo-Tiba Guiguisia, and M.~Beye for the numerous discussions during their internship, which inspired the author to pursue and complete this work. W.A.~acknowledges doctoral funding from Aix-Marseille Université. I.M.~conceived the idea and the plan for the project. W.A., I.M., S.S., and R.H. performed the numerical calculations, while I.M.~and P.L.~derived the analytical results. All authors were involved in the analysis of the results and the writing of the manuscript.

% %%%%%%%%%%%%%%%%%%%%%%%%%%%%%%%%%%%%%%%%%%%%%%%%%%
\section{Data Availability Statement}

    Data supporting the findings of this study are available from the corresponding author on a reasonable request.

\nocite{*}
\bibliographystyle{apsrev4-1}
%\bibliography{aipsamp}

\providecommand{\noopsort}[1]{}\providecommand{\singleletter}[1]{#1}%

\end{document}